# Que Bian : An Electronic Medical Record Management System on Blockchain

Hao Wang
Ratidar.com
Beijing, China
haow85@live.com

## ABSTRACT

Medical Record Management System is an important information management system in healthcare centers and hospitals. Information kept in such systems need to be clean , correct and tamper-proof. In this paper, we take advantage of blockchains' tamper-proof and decentralization properties to develop a robust and secure electronic medical record management system. In particular we choose HyperLedger Fabric as our underlying technical architecture. HyperLedger Fabric yields higher throughput and lower latency compared with other blockchains, which is a perfect candidate for enterprise software development. Our system is a novel innovation that can serve as an ideal replacement for conventional Medical Record Management System.

## CCS Concepts

• **Information systems** → Information Systems Applications

## Keywords

blockchain , HyperLedger , Medical Record Management System, Identity Access and Management System

## 1. Introduction

Blockchain is one of the most heavily invested technologies today. Since its invention in 2009, proof-of-concept work of blockchains has been widely developed in fields such as finance and taxation. Blockchain is essentially a distributed ledger where the transaction logs are tamper-proof and append-only. At its core, it borrows lavishly from distributed computing domain. As a distributed system, blockchain needs multiple nodes of the system to reach consensus in the process of computing. Design of consensus protocols is key to blockchain system. Different consensus protocols leads to different blockchain systems that vary dramatically in performance.

However, the major technical bottleneck of blockchains has been efficiency of consensus protocols. Old blockchain systems like Bitcoins and Ethereum suffer severely from its efficiency problems such as extremely low throughput and high latency. By late 2018, both Bitcoin and Ethereum could not generate TPS higher than 20. The low efficiency has given blockchain huge disadvantages compared with conventional centralized IT systems. Researchers and industrial engineers have thus devoted a huge amount of resources continuously working on improvement of blockchain efficiency with many new consensus protocols invented each year. Unluckily there is no "one-size-fits-all" consensus protocols that fits all kinds of application scenarios.

A particular useful design of blockchain HyperLedger is its consensus protocol is pluggable. HyperLedger is a distributed ledger developed by IBM. It is one of the most popular smart contract technical backends. HyperLedger is a permissioned blockchain with pluggable consensus protocols that yields thousands of TPS. It is a much faster and scalable blockchain than other technologies such as Ethereum. HyperLedger has an active community with contributors from both academia and industry. It is one of the most suitable technologies for enterprise software development.

Electronic medical record management system is an indispensable part of modern healthcare systems. Diagnostics of patients' symptoms need to be accurately documented in the system for physicians' reference. Data in electronic medical record should be clean, up-to-date and tamper-proof. Blockchains is an intrinsically fit technical choice for electronic medical record management systems.

Not only the document content of electronic medical record management system needs to be carefully kept. It is also crucially important to securely manage the identities of the patients of the electronic medical record management system. Blockchain in this case is also a perfect fit for the task. We create a blockchain- powered Identity and Access Management System (IAM) based on HyperLedger Indy to manage the patients' identities.

## 2. Related Work

Bitcoin [1] is the first real-world application of blockchain technology. Blockchain is a distributed ledger which allows users to record transactions in a distributed way.To guarantee security, Bitcoin provides incentives for users involved in the data validation process. Such incentives are called cryptocurrencies. They started to embody real-world monetary values in the past 10 years. In 2017, Bitcoins hit the historic peak value of 1 BTC equivalent to over 20K USD.

Following the invention of Bitcoin, other blockchain technologies have also emerged. Alongside the original ledger functionality, new blockchain systems provide other functionalities such as smart contracts. Two very popular blockchains which are widely used for their smart contract functionalities are Ethereum [2] and HyperLedger [3] . Ethereum uses PoW as its consensus protocol, which suffers from very low TPS. Most Ethereum applications are not comparable with conventional centralized IT systems when it comes to performance. HyperLedger is a blockchain project initiated by IBM. It now embodies several sub-projects such as Fabric, Indy etc. HyperLedger uses a pluggable consensus protocol which could generate thousands of TPS.

There have been tremendous amount of research work on consensus protocols. The most popular consensus protocols in modern days are Byzantine Fault Tolerance algorithms. It is well known that deterministic Byzantine Fault Tolerance algorithms do not have polynomial time solutions. Variants such as Practical Byzantine Fault Tolerance [4] and Redundant Byzantine Fault Tolerance [5] have been well integrated into commercial solutions. Since different application scenarios have different demands for blockchain systems, researchers have compared among different consensus protocols and proposed approaches where consensus protocols are switched adaptively.To provide more flexibility in consensus protocol usage, HyperLedger designs its consensus protocol schema as a pluggable one with which users could choose their own favorite protocols.

Identity and Access Management System (IAM) has bred a series of companies like ShoBadge. IAM system is also an inseparable part of our electronic medical management system where physicians' and patients' identities need to be kept clean and secure.

## 3. Blockchain

Most old-school blockchain systems are built upon the order- execute-validate paradigm. Blockchain is a distributed ledger composed of multiple peer nodes in the system. When transactions take place, they are sent to every node in the system. Then they get ordered sequentially by the peers they are sent to. A validation process follows where most consensus protocols take effect. After the validation process, the transactions are effectively executed in order. The entire computational process then completes. Blockchain systems such as Bitcoin and Ethereum all follow this paradigm.

Unlike the old order-execute-validate paradigm, HyperLedger invented a new system where transactions take effect. Transactions take place in the execute-order-validate manner. In this paradigm, execute, order and validate phases could be executed in parallel, thus greatly enhanced the computational efficiency of the blockchain system. Transactions are first sent out to peer nodes where they get endorsed, then they are ordered by the ordering services, in the last validate phase, transactions are checked for conflicts and thenexecuted.

In the order phase of HypeLedger, consensus protocols are used to reach agreement on how transactions should be ordered sequentially. HyperLedger allows users to choose the most suitable consensus protocol for the current application scenario. Potential candidates include PBFT (Practical Byzantine Fault Tolerance) and many other BFT (Byzantine Fault Tolerance) variants.

## 4. HyperLedger Fabric & Indy

HyperLedger Fabric is part of the HyperLedger suite. It uses Kafka cluster to implement its underlying consensus protocol. As a permissioned blockchain, it can be safely distributed and deployed among multiple nodes across different healthcare departments or organizations.

HyperLedger Indy is also a sub-project of HyperLedger. It is initiated by Evernym and developed by Sovrin. HyperLedger Indy is an identity management system built upon blockchain. It is well supported through blockchain technical foudations and communities. HyperLedger Indy is chosen as the backbone of our Identity and Access Management System due to its maturity and technical soundness.

## 5. Identity and Access Management System

An identity in our society has different embodiments: A person is a father in a family while at the same time he is a manager at a software company. He is a Wells Fargo bank account holder while at the same time he is a commuter of the Salt Lake City light trail system. Since different embodiments of the same identity are maintained and managed in heterogeneous IT systems, they are prone to nefarious behavior and mismanagement.

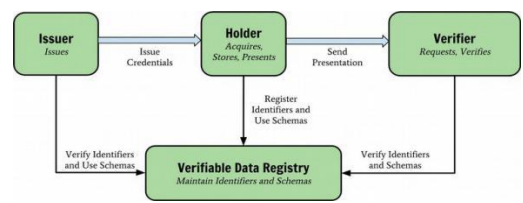

Fig 1. General Workflow of Hyperledger Indy

Due to its decentralization and tamper-proof properties, blockchain could provide better security for Identity and Access Management System (IAM) :

[1] Users control their own data without the risk of data leakage on third-party platforms.
[2] Multiple roles of the same identity could not be associated by public data.
[3] Authentication of identity by double factors or

multiplefactors.

In a decentralized IAM system, users store their identities by themselves, for example, on their own cell phones, therefore eliminating the risk of data leakage and mismanagement by centralized IT systems. We built our IAM system based on HyperLedger Indy as follows : The Identity Provider Services stores and publishes the identity data first, then it uploads the encrypted information associated with identities to the blockchain. The identity holder acquire the identity from the Identity Provider Services off-line and hold the identity. A service provider who needs the identity request the identity holder for an ID proof. The identity holder allows the service provider to verify his identity on blockchain. The service provider authenticates the ID on blockchain. After authentication, the entire workflow of blockchain IAM system finishes.

The main contributions of our IAM system compared with other systems are: 1. Our system is a blockchain powered system, which means it is secure , public and tamper-proof ; 2. Our system does not upload data onto the blockchain, which greatly reduces the risks of dataleakage.

## 6. Electronic Healthcare Management System

MIT media lab has been working on its own version of electronic healthcare records (EHR) system. The code named MedRec project [7] explores the opportunity to enhance human physical well-beings by taking advantage of the latest trend of technological innovation. It highlights the importance of data security in EHR systems.

MedRec is built upon Ethereum and fully utilizes its smart contract functionality to implement a full-featured electronic healthcare records system. The workflow diagram of MedRec is illustrated in Fig. 2.

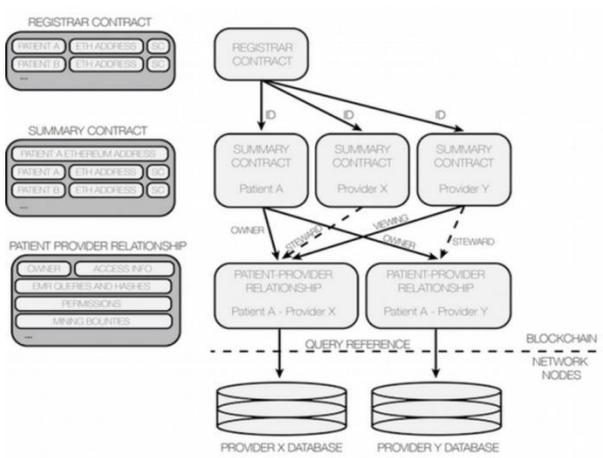

Fig. 2 Workflow Diagram of MedRec

## 7. System Design

We design our system upon HyperLedger Fabric and HyperLedger Indy as follows : we introduce 3 types of participants in our blockchain system, namely hospitals, doctors and patients. A hospital hosts doctors of different departments and patients affiliated with them. It has an ID, address, phone-number, department names, doctor ID's and patien tID's. A doctor has an ID, demographic information and the ID list of his patients. A patient has an ID, symptom ID and names, demographic information and other attributes.

The EHR system combines the IAM system with the health record management system built upon HyperLedger Fabric. The health record management system supports the following operations on data : Data queries by fields : Patient ID , Symptom ID , etc. ; Data append functionality which is the only write operation in the system. The entire system is succinct, both easy-to-implement and easy-to-use.

HyperLedger enables the separation of core technical functionality and product feature design and engineering. We built our non-blockchain part with Python , HTML, CSS and Javascript. The technologies and designs could be easily substituted with other alternatives without major modification to the blockchain backbone.

## 8. Conclusion

In this paper, we propose an EHR system called Que Bian. The system is built upon the blockchain HyperLedger Fabric and HyperLedger Indy. HyperLedger Fabric is used to implement the CRUD operations of the system, while HyperLedger Indy serves as the IAM system backbone. Our system is among its first kind in the medical field. It has far reaching outcomes in the incoming medical revolution involving new technologies such as artificial intelligence and blockchain.

In the future, more features will be added to our product and improvement over the computational efficiency will also be implemented. In specifics, the research focus will be optimization of underlying blockchain protocols and tuning of architecture and supportive technologies in the particular area of medical record management system.